\documentstyle[psfig]{mn}
\def\H0{{\it H}$_0$}
\def\Ms{{\it M}$_\odot$}

\def\q0{{\it q}$_0$}

\def\ergps{erg~s$^{-1}$}
\def\kmpspMpc{km~s$^{-1}$~Mpc$^{-1}$}
\def\Ms{{\it M}$_\odot$}

\def\min{$^{\prime}$}
\def\sec{$^{\prime\prime}$}
 
\def\psqcm{cm$^{-2}$}
\def\ergpspsqcm{erg~cm$^{-2}$~s$^{-1}$}

\def\cps{ct\thinspace s$^{-1}$}

\def\Rs{$R_{\rm S}$}

\def\rg{$r_{\rm g}$}

\title[X-ray periodicity in IRAS18325--5926] 
{Detection of an X-ray periodicity in the Seyfert galaxy IRAS~18325--5926} 
\author[K. Iwasawa et al] 
{\parbox[]{6.5in} {K. Iwasawa$^1$, A.C. Fabian$^1$,
W.N. Brandt$^2$, H. Kunieda$^3$, K. Misaki$^3$, C.S.~Reynolds$^4$ 
and Y.~Terashima$^3$}\\
\\
$^1$ Institute of Astronomy, Madingley Road, Cambridge CB3 0HA\\ 
$^2$ Department of Astronomy and Astrophysics, The Pennsylvania State
University, University Park PA 16802, USA\\ 
$^3$ Department of Astrophysics, Nagoya University, Chikusa-ku, Nagoya 464-01, Japan\\
$^4$ JILA, University of Colorado, Campus Box 440, Boulder CO 80309-0440, USA}
\date{}

\begin{document}

\maketitle

\begin{abstract}
We report the detection of a $5.8\times 10^4$ s periodicity in
the 0.5--10~keV X-ray light curve of the Seyfert galaxy
IRAS18325--5926, obtained from a 5-day ASCA
observation. Nearly 9 cycles of the periodic variation are seen; it
shows no strong energy dependence and has an amplitude of about 15 per
cent. Unlike most other well-studied Seyfert galaxies, there is no
evidence for strong power-law red noise in the X-ray power spectrum
of IRAS18325--5926. Scaling from the QPOs found in Galactic black hole
candidates suggests that the mass of the black hole in IRAS18325--5926
is $\sim 6\times 10^6$--$4\times 10^7$\Ms.
\end{abstract}


\section{Introduction}

ASCA X-ray spectra of many Seyfert galaxies show that the iron emission
line is broad and skewed to low energies (Tanaka et al 1995; Nandra et al
1997), indicating that the X-rays originate from above the surface of an
accretion disk close to a massive black hole (Fabian et al 1995).  The
line profiles demonstrate that much of the X-ray emission orginates from
about 10--20 gravitational radii, but do not constrain the mass of the
black hole. For that the emission radius is required in physical units.

X-ray variability is an important characteristic of such active galactic
nuclei (AGN) and may hold the key to obtaining that size. Short time
scale X-ray variations are probably due to individual flares above the
accretion disk and so do not determine the orbital radius unless the
flare mechanism is understood much more than is currently the case. A
periodic signal would suffice if the origin of the period, orbital or
otherwise, was known.

Previous studies of X-ray variability of AGNs have found that 
power spectra are featureless and
power-law in shape (e.g., Lawrence et al 1987; McHardy 1989). 
No characteristic periodicity
has been found except for possible quasi-periodic oscillations (QPOs) in
NGC5548 ($\sim 500 s$, Papadakis \& Lawrence 1993; Tagliaferri et al 1996
argue against the presence of QPO in those data), NGC4051 ($\sim 1$
hr, Papadakis \& Lawrence 1995) and RX\thinspace J0437.4--4711 
($0.906\pm 0.018$ dy, Halpern \& Marshall 1996). 
Very clear QPO have been seen in several
Galactic Black Hole Candidates; a 67 Hz QPO in GRS 1915+105 
(Morgan, Remillard \& Greiner 1997) and QPO at
lower frequencies, 1--10~Hz, in GX339-4 and GS~1124--68 (Dotani 1992;
Belloni et al 1997).

Here we report the discovery of a periodic signal from IRAS18325--5926,
which is a Seyfert galaxy with a broad iron line (Iwasawa et al 1995,
1996a). The X-ray source is one of the complete X-ray sample selected by
the HEAO-1 A2 survey (Piccinotti et al 1982) and it was identified with
the IRAS galaxy (= Fairall 49) by Ward et al (1988). Although its
narrow-emission-line-dominated optical spectrum is similar to that of
Seyfert 2 galaxies (DeGrijp et al 1985), the presence of a weak broad
wing to the Balmer lines (Carter et al 1984; Iwasawa et al 1995) suggests
a dust-obscured broad-line region in this galaxy. The X-ray spectrum is
moderately absorbed by a column density of $\sim 1\times 10^{22}$\psqcm,
which is consistent with the picture of an obscured Seyfert 1 nucleus.
Apart from the slightly steep continuum slope ($\Gamma\sim 2.1$), the
absorption-free hard X-ray emission ($>2$ keV) has similar properties to
those of Seyfert 1 galaxies. The X-ray source has been found to be highly
variable, as observed in other Seyfert 1 galaxies such as MCG--6-30-15.

A five-day long ASCA observation with a net exposure time of $\sim$ 200
ks has been carried out with the aim of investigating X-ray variability
in the source, including the iron line and underlying continuum. The
general spectral properties of the X-ray source are similar to those seen
before, and will be discussed in another paper. We report in this Letter
on a timing analysis of the ASCA data.

% lite curve

\begin{figure*}
%\centerline{\psfig{figure=s0lite.ps,width=0.8\textwidth,angle=270}}
\centerline{\psfig{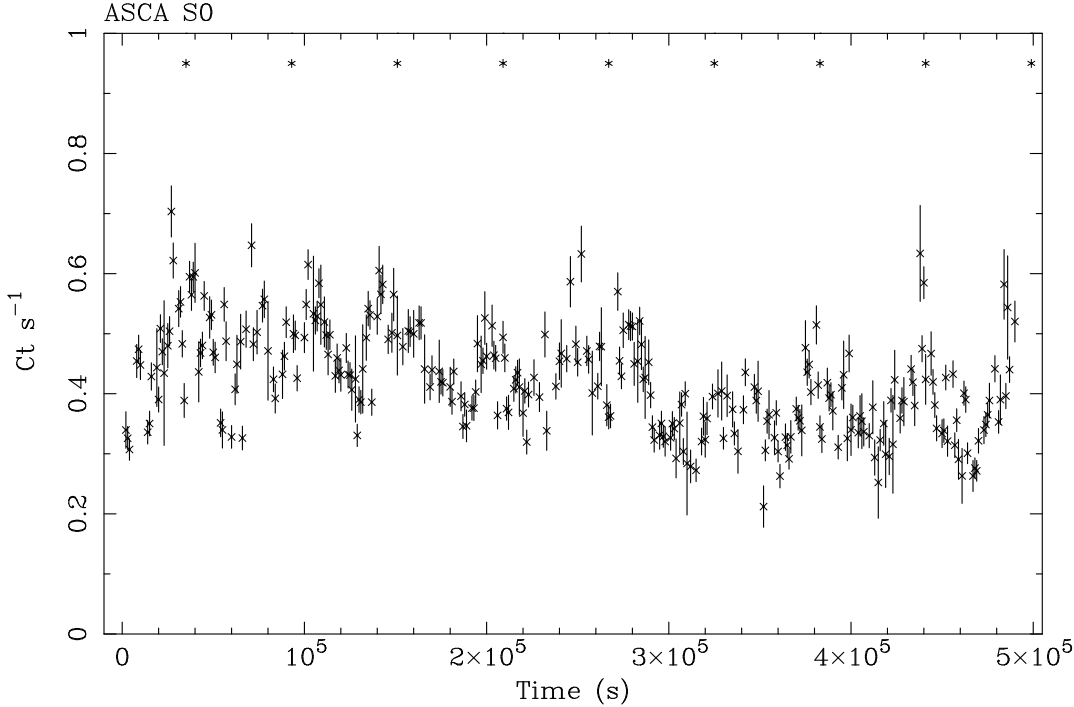}}
\caption{The ASCA S0 light curve in the 0.5--10 keV band. Each data bin
is of exposure time 1000 s for clarity while the timing analysis
used 360 s bins. The epoch of the start is 1997 March 27,
13:21:31 (UT). The nine stars plotted at the top are spaced at the
detected frequency.}
\end{figure*}

\section{The ASCA data}

IRAS18325--5926 was observed with ASCA for 5 days between 1997 March 27
and 1997 March 31. The X-ray telescopes pointed at the source at the 1CCD
mode nominal position. Averaged count rates detected with the four
detectors over the total observing run are shown in Table 1. The Solid
state Imaging Spectrometers (SIS; S0 and S1) are operated with the Faint
mode using one standard CCD chip in each detector. The standard settings
of PH mode were used for the Gas Imaging Spectrometers (GIS; G2 and G3).
The data reduction was performed using FTOOLS (version 4.0) and standard
calibration. After data selection, the net exposure times of useful
data are 204 ks for the SIS and 180 ks for the GIS.

The source events were collected from a circular region centred on the
source with a radius of 4 arcmin for the SIS and 6 arcmin for the GIS.
Background in each detector was taken from a source-free region on the
same detector. The backgrounds were basically stable with an intensity
about 7--8 per cent of the mean counts collected from the source regions.

The mean X-ray flux of the source is $1.7\times 10^{-11}$\ergpspsqcm ~in
the 2--10 keV band during the observation, a factor of $\sim 2$ brighter
than the previous ASCA observation in 1993 but a factor of $\sim 1.5$
fainter than the Ginga observation in 1989. The absorption-corrected
2--10 keV luminosity is $L_{\rm 2-10keV}\sim 3\times 10^{43}$\ergps
~($H_0 = 50$ \kmpspMpc).

The minimum time resolution of the data is 4 s. 
X-ray light curves from the four detectors have been
binned into 360 s for each data point for the timing analysis presented
below. The count rate decreased gradually over the whole observation with
many flares on short timescales (Fig. 1). The light curve appears to have
a characteristic time scale of $\sim 6\times 10^4$ s as well as random
shorter timescales. We have searched for a possible periodicity of the
X-ray modulation and examined its significance using a periodogram
technique.

%The X-ray flux of the source have changed over the long term:
%the observed 2--10 keV flux was $2.8\times 10^{-11}$\ergpspsqcm
%during the Ginga observation in 1989 (Iwasawa et al 1995) whilst
%it was $8.4\times 10^{-12}$\ergpspsqcm ~during the previous ASCA
%observation in 1993 September (Iwasawa et al 1996).

\begin{table}
\begin{center}
\caption{Mean count rates observed with each of the four detectors during the 
observation. No corrections for vignetting has been made.}
\begin{tabular}{cc}
Detector & Count rate \\
 & \cps \\[5pt]
S0 & 0.372 \\
S1 & 0.307 \\
G2 & 0.280 \\
G3 & 0.332 \\
\end{tabular}
\end{center}
\end{table}

\section{Timing analysis}

\begin{table}
\begin{center}
\caption{Results of the Lomb periodgram analysis. Probabilities that the
signals could occur randomly from the four detectors are shown for the
peak at $1.725\times 10^{-5}$ Hz. G2 (which had the lowest count rate from
the source) has a more significant peak at a frequency $1.27\times
10^{-3}$ Hz, not seen in the other detectors.}
\begin{tabular}{lc}
Detector & Probability \\[5pt]
%S0 & $2\times 10^{-11}$ \\
%S1 & $5\times 10^{-11}$ \\
%G2 & $6\times 10^{-4}$ \\
%G3 & $1\times 10^{-8}$ \\
S0 & $1\times 10^{-10}$ \\
S1 & $9\times 10^{-9}$ \\
G2 & $6\times 10^{-4}$ \\
G3 & $3\times 10^{-6}$ \\
\end{tabular}
\end{center}
\end{table}

The Lomb algorithm (Lomb 1976) is appropriate for searching for a
periodic modulation from unevenly sampled data such as inherent in
observations from low orbit satellites like ASCA (Tanaka, Inoue \& Holt
1994). In Fig. 2, the power-spectrum computed using the Lomb periodogram
(Press et al 1992) is shown.

\begin{figure}
\vspace{1cm}
\centerline{\psfig{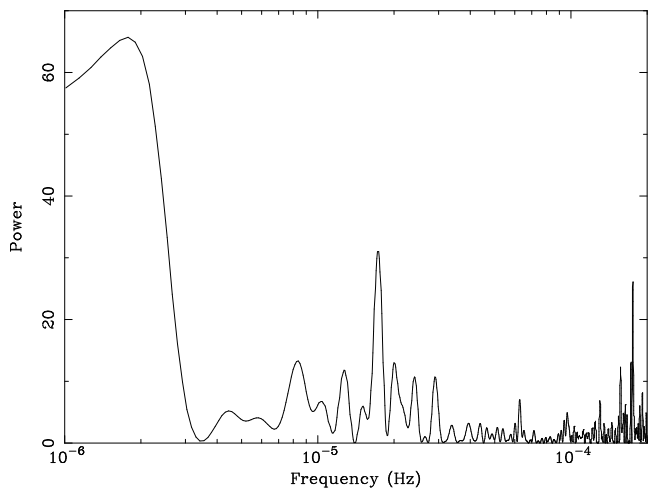}}
\caption{The power spectrum derived from the ASCA S0 light curve
of IRAS18325--5926 in the 0.5--10 keV band. A peak at the right
corresponds to the ASCA orbital period. The highest peak at $1.725\times
10^{-5}$ Hz is intrinsic to the source. Large power at the low frequency
below $3\times 10^{-6}$ Hz is due to the decreasing trend across the
whole observation. Note that a frequency corresponding to the
whole observation length is $\sim 2\times 10^{-6}$ Hz.}
\end{figure}

A large peak with a period just longer than the dataset and due to the
trend seen across the light curve, lies at the lowest frequency shown
in Fig. 2. 
There are two other significant peaks. The higher frequency peak at
$1.72\times 10^{-4}$ Hz or 96.9 min is likely to originate in the
orbital period of ASCA (Tanaka et al 1994). Another, more significant,
one is found at $1.725\times 10^{-5}$ Hz, corresponding to a period of
58.0 ks. This peak is found strongly in all detectors (Table
2) except for G2 in which the observed count rate is the lowest among
the four detectors thus the data are most noisy. The probabilities of
the peak being random are shown in Table 2. If there is significant
clumping in the data sampling, then the derivation of the probability
could be affected (Horne \& Baliunas 1986).  We have verified that
this is not the case using simulated light curves, one in which random
count-rates (with the same mean and variance as the data) are assigned
to the observed times and the other in which the count rates were shuffled.

The periodicity is confirmed with an epoch-folding technique.
The 90 per cent statistical error to the 58.0 ks
peak for the observed data in the $\chi^2$--(trial period) diagram,
when it is fitted with a gaussian, is $\sim 0.3$ ks. The systematic
error in deriving a coherent periodicity is found to be $\sim 0.2$ ks,
applying the same technique to simulated sine waves with a period of 58.0
ks, holding the same time locations and similar mean count-rate and
amplitude to the real observation. Thus the error of the period is
estimated to be $\sim 0.4$ ks.

The folded light curve at the detected period is shown in Fig.
3. The modulation is arch or sinusoidal in shape with an amplitude of 
$\sim 15$ per cent. There is no strong evidence for X-ray colour 
variation associated with the periodicity.

\begin{figure}
\vspace{5mm}
%\centerline{\psfig{figure=fold.ps,width=0.48\textwidth,angle=270}}
\centerline{\psfig{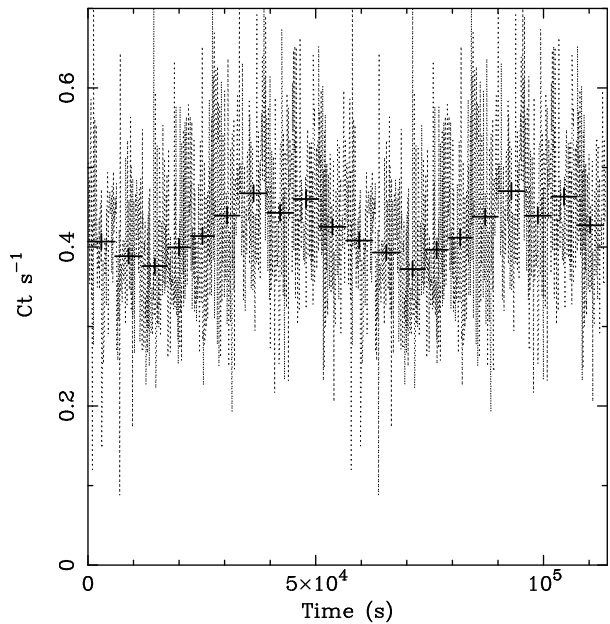}}
\caption{The ASCA S0 light curve folded at the detected period. 
Two cycles are shown for clarity. The original data (dotted line) are
averaged into 10 bins per cycle (thick solid line).
Error bars to count rates are calculated from the square
root of the number of counts integrated over each bin while
the standard deviation in each bin is typically $\sim 0.12$ \cps.}
\end{figure}

\begin{figure}
\vspace{-5pt}
\centerline{\psfig{figure=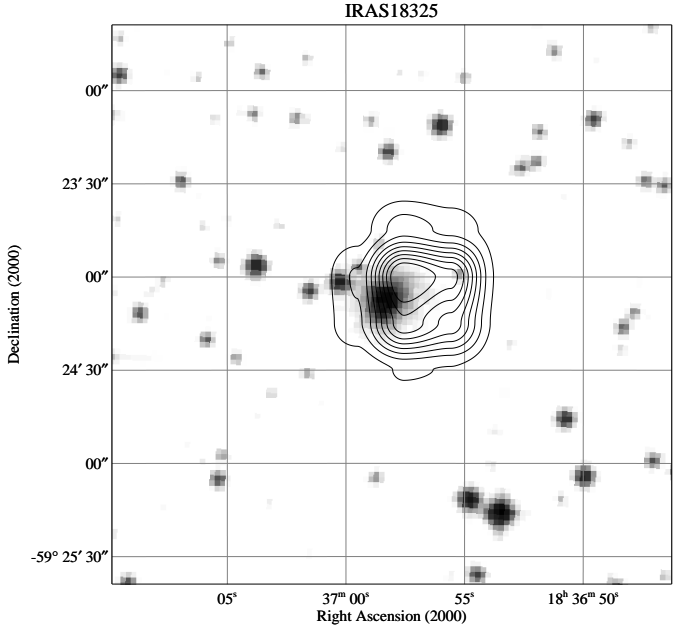,width=0.5\textwidth,angle=0}}
\caption{The ROSAT PSPC 0.5--2 keV contour image overlaid on the digitized
UK Schmidt survey plate of the IRAS18325--5926 region.}
\end{figure}

We note that the power spectra of the variability of Seyfert galaxies is
typically power-law in shape over the frequency intervals shown in Fig.~2
(McHardy 1989). The possibility of such red noise complicates the true
assignment of significance to the peak in Fig.~2. The probabilities in
Table~2 are strictly for a sinusoidal signal in white noise. Apart from a
large peak at the lowest frequencies due to the trend in the lightcurve,
we do however see no evidence for red noise. Over the frequency range
$3\times 10^{-6}$~Hz to above $10^{-4}$~Hz, the peaks in the power
spectrum are similar to those expected from white noise, apart from the
spike at $1.73\times 10^{-5}$~Hz. If it is argued that the spike is due
to red noise, then we must ask why there are no similar large peaks at
$10^{-5}$~Hz and below. We find that the power spectrum (and lightcurve)
are not at all similar in shape to simulations made with red noise. (The
light curve of MCG--6-30-15 from the previous ASCA long look ($\sim 4$
dy, Yaqoob et al 1997) does show red-noise.)

We made 1000 simulated lightcurves with the observed
time sequence and (phase) randomized $f^{-1}$ noise, normalized to give
the same variance as the real dataset. The minimum power (in the same
units as Fig. 2) in the frequency range from $3\times 10^{-6}$~Hz to
$1.5\times 10^{-5}$~Hz was 42, decreasing to 25 when the lower frequency
was $4\times 10^{-6}$~Hz. 
If the power varies as $f^{-1.5}$, which is more typical of AGN,
then these values rise to 80 and 60 respectively. 
All of these far exceed the power seen in that frequency range in the
real dataset. 
If the $f^{-1}$ power is scaled down to match the observations then the
maximum power seen in the simulation beyond $1.5\times 10^{-5}$ is less 
than 10. We conclude that the observations are inconsistent with
any simple power-law red noise model.

The peak at $1.725\times 10^{-5}$ Hz is nearly 10 times the frequency of
the peak due to the ASCA orbital period. We consider this to be a
chance coincidence. To the best of our knowledge, there is no
corresponding frequency originating in the spacecraft. This has been
checked using the light curves of the calibration source of the two
GIS detectors during the present observation and the light curve of 
two-day long ASCA observation of a
stable X-ray source, the Centaurus cluster, kindly provided by S. Allen.
An examination of the X-ray image rules out a possibility that wobbles of
the spacecraft causes the observed X-ray modulation.

Note that the X-ray source is the brightest both in the ASCA and ROSAT
PSPC images of the region and is located at the optical position of the
galaxy (18$^{\rm h}36^{\rm m}58^{\rm s}$, $-59^{\circ}$24\min 09\sec,
J2000). Simultaneous ROSAT and ASCA observations of the source in 1993
(Iwasawa et al 1996) show a correlated decrease in flux by a factor of
about two, confirming that the ROSAT and ASCA sources are the same. We
show in Fig. 4, the ROSAT PSPC image overlaid on the digitized UKSTU
optical image. The offset of the X-ray and optical position is $\sim 10$
arcsec, which is within the PSPC positional uncertainty.

To eliminate further any residual doubt that the variability reported
here is from IRAS18325--5926, since the period is in the range of that of
low-mass X-ray binaries and some cataclysmic variables, we have estimated
the probability of a chance alignment with a Galactic object by scaling
from the Piccinotti et al (1982) X-ray sample. This implies about 100
Galactic objects in the Sky above $|b|=20^{\circ}$ at the flux level of
IRAS18325--5926 and thus probabilities of about $3\times 10^{-6}$ and
$10^{-7}$ of a further object in the 1 arcmin and 10 arcsec radius error
boxes obtained from the ASCA and ROSAT images, respectively. We note that
both images are consistent with a single point source. As mentioned in
the Introduction, the observed X-ray absorption is fully consistent with
the optical classification of the Seyfert galaxy, but unlikely for a
Galactic source. We conclude that it is most unlikely that we are
repeating the case of NGC6814, where the X-ray variability detected by
non-imaging detectors was due to a cataclysmic variable 40 arcmin from
the galaxy (Madejski et al 1993). The IRAS galaxy is by far the most
probable X-ray source.

\section{DISCUSSION}

The periodicity found in the light curve from the long observation is
about 58 ks which is probably intrinsic to the source whilst the 97 min
peak is due to the ASCA orbital period. A similar X-ray flux variation is
apparent in the previous ASCA PV (Iwasawa et al 1996a) and Ginga (Iwasawa
et al 1995; Smith \& Done 1996) observations. 
The duration of each of those observations was
about one day and so only slightly longer than the period ($\sim$16 hr).
The data sets are too widely spaced to phase together.

The (Newtonian) orbital period at 5\Rs ~= 10\rg ~is $3.4\times 10^{4} M_8
R_1$ s where the mass is in $10^8 M_8$\Ms ~and the radius is $10 R_1$\rg.
So it could easily be occurring at 10\rg ~for a $1.7\times 10^8$\Ms
~black hole or at 20\rg ~for a $2.1\times 10^7$\Ms ~one. In other words,
a 16 hr period is reasonable for orbital variations at 10--20
gravitational radii in a disk around a black hole of mass $2\times 10^7 -
10^8$\Ms. The Eddington limit for such masses is above $10^{45}$\ergps,
well above that observed.

What we do not know from the 9 cycles that we have seen is whether the
variation is strictly- or quasi-periodic. Further observations are
necessary to test whether the observed periodicity is long-term or part
of a QPO. The quality factor of the current variation is at least 30 and
could be much higher. A strict periodicity will be difficult to explain.
The most obvious clock would be an object in orbit about the accreting
black hole. The lifetime against spiral-in due to gravitational radiation
losses would be only weeks for another black hole of similar mass to the
primary (say $10^7$\Ms), increasing as the inverse mass for much
smaller objects. A star orbiting within an accretion disk does isolate a
particular radius (see discussion by Syer, Clarke \& Rees 1991) and could
last 100,000~yr but it is not obvious how it could modulate the X-ray
emission.

A transient periodicity could be due to a very long-lived flare, or flare
patch, on the disk. The appearance of small variation originating at a
few Schwartzschild radii may be boosted relativistically by orbital motion
(see e.g., Boller et al 1997).

A 67 Hz QPO has recently been discovered with the Rossi X-ray Timing
Expolorer (RXTE) in the Galactic black hole candidate GRS 1915+105
(Morgan, Remillard \& Greiner 1997). If this and the periodicity seen in
IRAS18325--5926 are related to the orbital period at the same radius (in
gravitational units) in each system, then the masses of their black holes
should scale as $67/1.73\times 10^{-5}$ which is about 4 million. Assuming
that the black hole in GRS 1915+105 is 10\Ms\ then we find that the mass
of the black hole in IRAS18325--5926 is $\sim 4\times 10^7$\Ms. We note
that both GRS 1915+105 and IRAS18325--5926 are observed at moderately
high inclinations, of about 70 deg and 50 deg, respectively.

QPO at lower frequencies, 1--10~Hz, have been seen in the Galactic Black
Hole candidates GX339-4 and GS~1124--68 (Dotani 1992; Belloni et al 1997). 
The origin is again unknown, but simple scaling would then indicate a mass
of $4\times 10^6$\Ms, dependent on the black hole mass of the Galactic
sources. In these cases, the QPO are seen mostly when the source is very
luminous, possibly close to the Eddington limit, although they were also
seen in GS~1124--68 four months after its peak. 

It is unclear why any particular radius is favoured. One possibility is
the disk oscillation model proposed by Nowak et al (1997) for GRS
1915+10. General relativistic effects cause the epicyclic frequency for
matter in the disk to have a maximum close to, but not at, the inner
radius of the disk. This can lead to g-mode oscillations in the disk
causing QPO at that frequency. However, as discussed by Nowak et al
(1997) the power in such QPO should only be a few per cent of that of the
disk emission. It is unlikely that they cause a 10 per cent oscillation
of the emission from the corona above the disk.

Perhaps the spin of the black hole is responsible, if its axis is offset
from that of the disk. The Bardeen-Petterson effect (1975) due to the
dragging of inertial frames close to the black hole then causes the
inclination of the disk to change at a radius of some tens of
gravitational radii (see Rees 1984). If the inclination of the disk is
moderately high then oscillations near that radius may be detectable
through obscuration and/or reflection.

\section*{Acknowledgements}

We thank all the member of the ASCA team. T. Dotani is thanked for helpful
discussion. 
The ROSAT PSPC image was retrieved from the ROSAT archive maintained at
the Goddard Space Flight Center.
The optical UK Schmidt image was taken from the Digitized Sky Survey
produced by Space Telescope Science Institute and the original data
were taken by Royal Observatory Edinburgh.
ACF and KI thank Royal Society
and PPARC, respectively, for support.
CSR thanks the National Science Foundation for
support under grant AST-9529175.

\end{document}